\documentclass{ceab}   %% loading ceab.cls creates CEAB style

\usepackage{epsfig}     %\  Used to include figures
\usepackage{graphicx}   %/

\usepackage{ceabbib}     % for producing References section 
\usepackage[T1]{fontenc}% for producing Polish vowels 

\begin{document}

\title{RHESSI investigation of X-ray coronal sources during decay phase of solar flares: II. energy balance}

\author{S. KO{\L}OMA\'NSKI, Z. KO{\L}TUN, T. MROZEK, U. B\k{A}K-ST\k{E}\'SLICKA
\vspace{2mm}\\
\it Astronomical Institute of the University of Wroc{\l}aw,\\
\it Kopernika 11, 51-622 Wroc{\l}aw, Poland}

\def\gore{}

\maketitle

\begin{abstract}
As it was shown by many authors, a slow decrease in X-rays observed during the decay phase of long duration flares (LDE) can be explained only by a magnetic reconnection and energy release ceaselessly ongoing in the coronal part of a flare. Using \textit{RHESSI} data we try to answer two following questions. How effective are these processes at the LDEs decay phase and how can precisely the energy release rate be calculated based on these data? To answer the questions images of the selected LDEs during their decay phase were reconstructed. Physical parameters of flare coronal sources obtained from image spectral analysis allowed us to study the efficiency of the energy release process. We also examined terms included in the energy equation to find out what is the accuracy of determination of each term.
\end{abstract}


\section{Introduction}

A long duration event (LDE) is a solar flare characterized by a slow decrease in soft X-ray (SXR) emission. This decrease may last many hours.
Much insight into the nature of LDEs was made by ultraviolet and X-ray observations, during \textit{Skylab}, \textit{SMM} (Solar Maximum Mission) and \textit{Yohkoh} space missions (e.g. Sheeley et al., 1975; Kahler, 1977; Feldman et al., 1995; Harra-Murnion et al., 1998; Czaykowska et al., 1999; Shibasaki, 2002; Isobe et al., 2002). One of the most important conclusions is that without the continuous energy input during the whole decay phase LDEs would decay much faster than it is observed.

Loop-top sources (LTSs) are remarkable SXR and HXR (hard X-rays) features of solar flares seen close to a flare loop apex. They form before flare maximum and in LDEs may last whole decay phase (e.g. Feldman et al., 1995; Ko\l oma\'nski, 2007a). LTSs were first recorded on images taken from \textit{Skylab} and are commonly present in the present-day flare observations. The sources should be located close to the primary energy release site (e.g. Kopp \& Pneuman, 1976; Shibata, 1999; Hirose et al., 2001; Karlick\'{y} \& B\'arta, 2006). This fact makes them a very promising phenomenon for the analysis of the energy release during the decay phase. Since the first observation it has become clear that presence of LTS during the whole flare decay-phase requires continuous energy release and some restriction mechanism efficiently preventing outflow of mass and energy from LTSs (see Vorpahl et al., 1977). Without meeting these two requirements loop-top sources would rapidly lose energy by radiative and conductive processes and would vanish. This result was later confirmed by an analysis based on \textit{Yohkoh} and \textit{RHESSI} data (e.g. Jiang et al., 2006; Ko\l oma\'nski, 2007b).

An analysis of energy release during the decay phase can give us precise constraints for numerical flare models. The most demanding for the models is long-lasting HXR emission of LTSs seen in LDE flares. Therefore such LTSs observed for many hours after flare maximum are probably the most promising observational feature to set up the constraints. The sources should be located very close to the energy release site and, that is the most important attribute, they put high requirements on the energy release rate. If a long-lasting HXR source is thermal then it must be continuously heated, because the characteristic radiative cooling time of hot (above 10 MK) and dense ($\approx10^{10}\;{\rm cm}^{-3}$) plasma is about 1 hour. If an HXR source is non-thermal then there should be a continuous acceleration of particles, to counteract fast (about several seconds) thermalization of non-thermal electrons.

Here we present the investigation of energy release in nine LDEs observed by {\it RHESSI}. The analysis is made using {\it RHESSI} images reconstructed in narrow ($1$~keV) energy intervals. Our results are presented in to papers. In the first part (Mrozek et al. 2011, this issue, hereafter Paper~1) we showed image reconstruction technique and estimation of LTSs physical parameters and size through imaging spectroscopy. In this second part we use the parameters to calculate the energy balance of the observed sources and to find out how effective are the energy release and heating processes at the decay phase of LDEs.

\section{Observations}

For our analysis we selected nine LDEs well observed by \textit{RHESSI} (Lin et al., 2002). We chose flares of significantly different power and with 
decay phases lasting more than 7 hours in \textit{GOES} $1-8$~\AA\,range. The flares listed in Table~1 in Paper~1. {\it RHESSI} data were supported with {\it SoHO}/EIT (Delaboudini\`ere et al., 1995) and {\it GOES}/SEM (Space Environment Monitor, Donnelly et al., 1977) observations. More information about the selected LDEs is given in Paper~1.

\section{Analysis}

\subsection{Geometrical and physical parameters of LTSs}

To estimate the heating rate of an LTS we need to know its geometrical (size, altitude) and physical (temperature, emission measure) parameters. These were determined from \textit{RHESSI} images. Image reconstruction, imaging spectroscopy, determination of size and physical parameters were described in Paper~1.

Estimation of the altitude of an LTS requires its position in the {\it RHESSI} image and determination of the point on the solar photosphere above which the source is situated (reference point). The source position was defined by the position of its centroid (see Paper~1). The position of reference point was taken from locations of flare ribbons recorded by \textit{SOHO}/EIT or determined using the method described by Roy \& Datlowe (1975). The method allow to estimate heliographic coordinates for flares behind the solar limb. We plot a position (on the solar disk) of an active region in which an analysed behind-the-limb flare occurred as a function of time. The position is taken from positions of all on-disk flares which occurred in that active region. Then we extrapolate the position vs. time plot behind the limb to get a position of our behind-the-limb flare.

All positions of reference points were corrected to account for the solar rotation. Finally, the LTS altitude was calculated as the distance between its centroid and the reference point. The altitudes obtained were corrected for projection effects. Errors in the location of the reference point and the location of source centroid were included in the altitude errors.

\subsection{Energy balance}

As mentioned in the Introduction, the presence of HXR emission from an LTS during the decay phase is evidence for energy release at that time. To calculate the heating rate of an LTS we considered its energy balance during the decay phase. Change of thermal energy of a loop-top source is due to some processes that cool and heat plasma of the source. Three major cooling processes where included into this balance: expansion, radiation and conduction. Knowing the change of LTS thermal energy and values of the three cooling processes we can calculate if and how efficiently the LTS was heated. The equation of energy balance can be written as follows:

\begin{eqnarray*}
\label{e0}
observed~change~of~thermal~energy = adiabatic~expansion - \cdots\\
- conductive~cooling - radiative~cooling + heating~rate
\end{eqnarray*}

or written in explicit form:

\begin{equation}
\label{e1}
\left (\frac{d{\mathcal E}}{dt}\right )_{obs} = \left( \frac{d{\mathcal E}}{dt} \right)_{ad} - E_C - E_R + E_H
\end{equation}

where: 

\begin{itemize}
	\item ${\mathcal E} = 3NkT$ is thermal energy density,
	\item $\left (\frac{d{\mathcal E}}{dt}\right )_{obs}$ is the decrease of ${\mathcal E}$ per second estimated from temperature ($T$) and density (number density of electrons, $N$) values,
	\item $\left (\frac{d{\mathcal E}}{dt}\right )_{ad}$ is the decrease due to the adiabatic expansion of plasma in a source,
	\item $E_C$ is the energy loss due to thermal conduction,
	\item $E_R$ is the radiative loss, and
	\item $E_H$ is the heating rate or thermal energy release.
\end{itemize}

The values of $E_C$, $E_R$ and $E_H$ are in erg~cm$^{-3}$~s$^{-1}$. We calculated:

\begin{itemize}
	\item $\left (\frac{d{\mathcal E}}{dt}\right )_{ad} = 5kT\left (\frac{dN}{dt}\right)$,
	\item $E_C = 3.9\times10^{-7}T^{3.5}/(Lr)$ where $r$ is the LTS radius and $L$ is loop semi-length (Jakimiec et al., 1997), and
	\item $E_R = N^{2}\Phi(T)$ ($\Phi(T)$ is the radiative loss function taken from Dere et al. (2009)).
\end{itemize}

We took the altitude of an LTS above the photosphere ($h$) as an approximation for $L$ in the expression for $E_C$. Of course, $h$ is smaller than $L$, but they do not differ too much (see subsection 'Accuracy of the energy balance terms -- conduction term').

\subsubsection{accuracy of the energy balance terms -- radiation term}

The radiative cooling term involves the radiative loss function. How precisely is this function known? The are many determinations of the radiative loss function in the literature (e.g. Cox \& Tucker, 1969; Raymond et al., 1976; Landi \& Landini, 1999; Reeves \& Warren, 2002; Colgan et al., 2008; Dere et al., 2009). They differ in adopted atomic data, set of elements and their abundances. These differences lead to discrepancies between the determined function. We are especially interested in the part of the function around 10~MK, because that is the typical temperature of LTSs. After inspection of the most recent determinations of the radiative loss function we can conclude that they do not differ more than by factor 3. Moreover the function around 10~MK is hardly sensitive to temperature change therefore errors in temperature do not cause large uncertainty in $\Phi(T)$. All in all radiative loss function is not the source of substantial errors in the radiative cooling term.

The term involves also three observational parameters: temperature, emission measure and size of an LTS. The last two are needed to calculate density. Temperature and emission measure are determined with quite a good accuracy (relative errors less than 10\%, see Paper~1) while LTS size is not (relative error usually greater than 10\%). Moreover, due to the method we used to reconstruct images (see Paper~1), there is possibility that size of LTS is overestimated. In such a case density and, in consequence, the radiative cooling term would be underestimated.

\subsubsection{accuracy of the energy balance terms -- conduction term} 

Thermal conduction term involves temperature, altitude and size of an LTS. Two first parameters are determined with quite a good accuracy (errors less than 10\%). Size, as mentioned, is the most uncertain parameter obtained from \textit{RHESSI} data. Fortunately, the term is dominated by temperature (in high power of 3.5) thus, it can be calculated with a good accuracy.

However, the equation we use for conductive cooling is valid for Spitzer conductivity, while the real conductivity in the solar corona can be suppressed due to the following factors:
\begin{enumerate}
	\item \textit{Non-local conduction.} It has been shown that if the temperature variation length scale is less than 30 times longer than the mean free path of the thermal electrons then the actual conductivity becomes smaller than Spitzer conductivity (Luciani et al., 1983). This is the case for a typical solar flare. For our LDEs flares the lowered non-local conduction may be 10 times smaller than Spitzer one.
	\item \textit{Suppressed outflow of mass and energy from LTSs.} Observations of LTS suggest that some kind of restriction efficiently preventing outflow of mass and energy even for hours must be present at the boundary of the sources (Vorpahl et al., 1977; Jakimiec et al., 1998; Jiang et al., 2006). In the extreme case the suppressed conduction flux may be even zero, but this scenario is rather too extreme.
\end{enumerate}

Taking these two factors into account we decided to calculate upper and lower limits for $E_H$. The upper limit, $(E_H)_{max}$, is calculated directly from Equation~(\ref{e1}) i.e. with Spitzer conductivity. As mentioned we took LTS altitude $h$ as an approximation for loop semi-length $L$. Because $h$ is always smaller than $L$, we slightly overestimate $E_C$ and in consequence $(E_H)_{max}$ can be regarded as unattainable upper limit. A lower (the lowest possible) limit, $(E_H)_{min}$, can be obtained assuming $E_C = 0$, though this may not be physically realistic. The actual value of LTS heating rate should be contained between the upper and lower limits of $E_H$.

\subsubsection{accuracy of the energy balance terms -- thermal energy and adiabatic expansion terms}

Thermal energy and adiabatic expansion terms involve the same set of observational parameters as the radiation term. Thus, there is similar problem with accuracy of these two terms caused by uncertainty of determination of an LTS size. Fortunately, both terms are small in value, usually smaller than thermal conduction and radiation terms. This fact results from slow temporal changes of LTS temperature and density (see Paper~1). In consequence uncertainty of thermal energy and adiabatic expansion terms is not a source of substantial errors in $E_H$.

\section{Results}

\begin{figure}[!ht]
 \begin{center}
   \epsfig{file=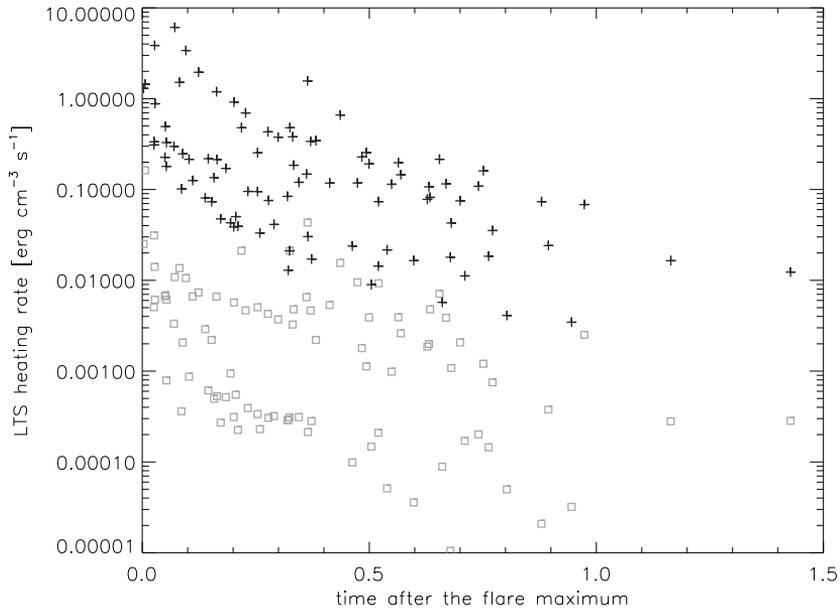,width=11cm}
 \end{center}
 \caption{Upper (pluses) and lower (squares) limits of the heating rate estimated for the nine analysed LDEs. Time after the maximum of a given LDE is expressed in units of characteristic time of temperature decay in this LDE.}
 \label{heating-rates}
\end{figure}

Estimated values of heating rates $(E_H)_{min}$ and $(E_H)_{max}$ for all analysed LDEs are shown in Fig.~\ref{heating-rates}. The figure shows how both limits of $E_H$ changed during the decay phase of a given LDE. Because the selected LDEs had different duration of the decay phase (see Table~1 in Paper~1) we decided to express time after a given LDE maximum in units of characteristic time of temperature decay in this LDE. Such an approach allows us to compare heating rates of all selected LDEs and to find out if there is any similar temporal behaviour of $E_H$ for LDEs of significantly different power and decay time. Values of heating rates $(E_H)_{min}$ and $(E_H)_{max}$ form quite clear pattern in the figure. There are two distinct paths which decrease slowly with time (lower path formed by $(E_H)_{min}$ and higher by $(E_H)_{max}$). The width of each path is not greater than 2 orders of magnitude. We observed that stronger flares do not lie higher in a path.

As one can see $E_H$ was always greater than zero, though the lower limit can be as small as $10^{-4}\;{\rm erg}\:{\rm cm}^{-3}\:{\rm s}^{-1}$. Even with such a low heating rate an LTS of typical volume of the of $10^{28}\;{\rm cm^{3}}$ existing for 10 hours needs in total more than $10^{28}\;{\rm erg}$ of thermal energy to be visible as it is. As we mentioned the $(E_H)_{min}$ may be always lower than actual heating. Assuming that actual $E_H$ is smaller than $(E_H)_{max}$ by a factor of 50 (we assumed that conduction is somewhat lowered by a presence of non-local conduction and by suppression of outflow of mass and energy from LTSs) we obtain that total energy needed to sustain typical long-lasting LTS during the decay phase is as high as $10^{31}\;{\rm erg}$. This is huge amount of energy which can be higher than energy released during the rise phase of an LDEs. Recently Jiang et al. (2006) investigated heating and cooling processes of an LTS and found that there is a large amount of energy released during the decay phase of a short-duration solar flare and that the amount may by comparable to amount of energy released during the impulsive phase. The authors suggest that in the case of LDEs the energy released during the decay phase may be even larger than the energy released during the impulsive phase. Our analysis may confirm this suggestion.

\section{Conclusions}

In several earlier papers authors estimated the rate of energy release in LDEs using the energy balance and data from \textit{Skylab} or \textit{Yohkoh} (e.g. Cheng, 1977; Isobe et al. 2002; Ko{\l}oma\'nski, 2007). \textit{RHESSI} data have several advantages comparing to previous instruments. {\it RHESSI} is very useful for analysing weak sources even with 1~keV energy resolution due to its high sensitivity. Owing to this we are able to perform precise imaging spectral analysis of LDEs. Moreover good spatial and energy resolution of the instrument, its high sensitivity to X-ray flux and to hot plasma allow us to obtain reliable values of all parameters needed for the energy balance calculations. As we showed, with the \textit{RHESSI} data we are able to calculate all terms of the energy balance equation with accuracy good enough to obtain reliable limits on the actual value of the heating rate:

\begin{itemize}
	\item The upper limit of the heating rate $(E_H)_{max}$ (with Spitzer conductivity), which is controlled mainly by the thermal conduction term, is determined with good accuracy.
	\item The lowest limit of the heating rate $(E_H)_{min}$ (with thermal conduction suppressed to zero), which is controlled mainly by radiation term, may by slightly underestimated due to overestimation of an LTS size.
\end{itemize}

Using the energy balance we estimated the heating rate for LTSs of nine LDEs during the decay phase. Our results can be summarized as follows:

\begin{itemize}
	\item The heating rate is non-zero during the whole decay phase and it decreases very slowly with time. This results in slow evolution of LTSs, e.g. very long characteristic time of temperature decay, and gives the sources their long-lasting existence.
	\item The total energy needed to sustain a typical long-lasting LTS during the decay phase can be as high as $10^{31}\;{\rm erg}$. This amount of energy can be equal to or higher than energy released during the rise phase of an LDEs. This severe requirement should be taken into account when building a model of solar flares.
\end{itemize}

\section*{Acknowledgements} 
The {\it RHESSI} satellite is NASA Small Explorer (SMEX) mission. We acknowledge many useful and 
inspiring discussions of Professor Micha\l \mbox{ }Tomczak. We also thank Barbara Cader-Sroka for 
editorial remarks. This investigation has been supported by a Polish Ministry of Science and High 
Education, grant No. N203 1937 33.
%\newpage

%%%%%%%%%%%%%%%%%%%%%%%%%%%%%%%%%%%%%%
% References produced by itemized list.
\section*{References}
\begin{itemize}
\small
\itemsep -2pt
\itemindent -20pt

\item[] Cheng, C.-C. 1977, \solphys, 55, 413

\item[] Colgan, J., Abdallah, J. Jr., Sherrill, M. E., et al. 2008, 689, 585

\item[] Cox, D. P., \& Tucker, W. H. 1969, \apj, 157, 1157

\item[] Czaykowska, A., De Pontieu, B., Alexander, D., \& Rank, G. 1999, \apjl, 521, 75

\item[] Delaboudini\`ere, J.-P., Artzner, G. E., Brunaud, J., et al. 1995, \solphys, 162, 291

\item[] Dere, K. P., Landi, E., Young P. R., et al. 2009, \aap, 498, 915

\item[] Donnelly, R. F., Grubb, R. N., Cowley, F. C. 1977, NOAA Tech. Memo. ERL SEL-48

\item[] Feldman, U., Seely, J.F., Doschek, G.A., \& Brown, C.M. 1995, \apj, {446}, 860

\item[] Harra-Murnion, L.K., Schmieder, B., van Driel-Gesztelyi, L., et al. 1998, \aap, {337}, 911

\item[] Hirose, S., Uchida, Y., Uemura, S., Yamaguchi, T., \& Cable, S.B. 2001, \apj, {551}, 586

\item[] Isobe, H., Yokoyama, T., Shimojo, M., et al. 2002, \apj, {566}, 528

\item[] Jakimiec, J., Tomczak, M., Fludra, A., \& Falewicz, R. 1997, Adv. in Space Res., 20, 2341

\item[] Jakimiec, J., Tomczak, M., Falewicz, R., Phillips, K.J.H., \& Fludra, A. 1998, \aap, 334, 1112

\item[] Jiang, Y.W., Liu, S., Liu, W., \& Petrosian, V. 2006, \apj, 638, 1140

\item[] Kahler, S. 1977, \apj, {214}, 891

\item[] Karlick\'{y}, M., \& B\'arta, M. 2006, \apj, 647, 1472

\item[] Ko\l oma\'nski, S. 2007a, \aap, {465}, 1021

\item[] Ko\l oma\'nski, S. 2007b, \aap, {465}, 1035

\item[] Kopp, R.A., \& Pneuman, G.W. 1976, \solphys, { 50}, 85

\item[] Landi, E. \& Landini, M. 1999, \aap, 347, 401

\item[] Lin, R.P., Dennis, B.R., Hurford, G.J., Smith, D.M., Zehnder, A., et al. 2002, \solphys, 210, 3

\item[] Luciani, J.F., Mora, P., \& Virmont, J. 1983, Phys. Rev. Lett., 51, 1664

\item[] Mrozek, T., Ko{\l}tun, Z., Ko{\l}oma\'nski, S., B\k{a}k-St\k{e}\'slicka, U. 2011, CEAB, this issue

\item[] Sheeley, N.R., Bohlin, J.D., Brueckner, G.E., et al. 1975, \solphys, 45, 377

\item[] Shibasaki, K. 2002, \apjl, {567}, L85

\item[] Shibata, K. 1999, \apss, 264, 129

\item[] Raymond, J. C., Cox, D. P. \& Smith, B. W. 1976, \apj, 204, 290

\item[] Reeves, K. K. \& Warren, H. P. 2002, \apj, 578, 590

\item[] Roy, J.-R., \&  Datlowe, D. W. 1975, \solphys, 40, 165

\item[] Vorpahl, J. A., Tandberg-Hanssen, E., \& Smith, J. B. 1977, \apj, 212, 550

\end{itemize}
%%%%%%%%%%%%%%%%%%%%%%%%%%%%%%%%%%%%

\end{document}